\documentclass[aps,prb,showpacs,superscriptaddress,twocolumn]{revtex4}
\usepackage{amsfonts}
\usepackage{subfigure}
\usepackage{txfonts}
\usepackage{amssymb}
\usepackage{amsbsy} 
\usepackage{epsfig}

\def\Re{\rm{Re}}

\def\be{\begin{equation}} \def\ee{\end{equation}}
\def\bea{\begin{eqnarray}} \def\eea{\end{eqnarray}}

\def\nn{\nonumber}

\def\bq{{\bf q}}

\def\bk{{\bf k}}

\def\be{{\bf e}}

\def\rw{\rightarrow}

\begin{document}

\title{ Collective modes in nodal line semimetals}

\author{Zhongbo Yan}
\affiliation{ Institute for Advanced Study, Tsinghua University,
Beijing,  China, 100084}

\author{Peng-Wei Huang}
\affiliation{ Institute for
Advanced Study, Tsinghua University, Beijing,  China, 100084}

\author{Zhong Wang}
\altaffiliation{ wangzhongemail@tsinghua.edu.cn} \affiliation{ Institute for
Advanced Study, Tsinghua University, Beijing,  China, 100084}

\affiliation{Collaborative Innovation Center of Quantum Matter, Beijing 100871, China }

\date{ \today}

\begin{abstract}

Recently, the nodal line semimetals have attracted considerable interests in condensed matter physics. We show that their distinct band structure can be detected by measuring the collective modes. In particular, we find that the dependence of the plasmon frequency $\omega_p$ on the electron density $n$ follows a $\omega_p \sim n^{1/4}$ law in the long wavelength limit. Our results will be useful in the ongoing search for new candidates of nodal line semimetals.

\end{abstract}

\pacs{73.43.-f,71.70.Ej,75.70.Tj}

\maketitle

\section{Introduction}

During the last decade topological insulators have generated intense
interests in condensed matter physics\cite{qi2010a,hasan2010,qi2011}.
These novel classes of material have insulating bulk but robustly
metallic surface. This peculiar feature has its origin in the
topological nontriviality of the band structures, described in terms of topological invariants\cite{thouless1982,niu1985,kane2005b,fu2007b,moore2007,qi2008,schnyder2008,wang2012a,
Chiu2015RMP,Wang2014ground-state}. Since the bulk samples of topological insulators are quite inert, all the highly interesting phenomena are produced by the topological surface states.

More recently, the topological properties of several classes of semimetals are recognized and discovered experimentally, among which are the topological Dirac semimetals\cite{liu2014discovery,neupane2014,Borisenko2014,xu2015observation, wang2012dirac,young2012dirac,wang2013three,Sekine2014,Zhang2015detection,
Chen2015Magnetoinfrared,Yuan2015ZrTe5} and Weyl semimetals\cite{wan2011,NIELSEN1981,nielsen1983adler,volovik2003,burkov2011}. In the Weyl semimetals, there exist topologically protected band touching points, termed as the Weyl points, which serve as magnetic monopoles of the Berry gauge field in the reciprocal space.  They exhibit novel transport and optical phenomena induced by chiral anomaly \cite{son2012,liu2012,aji2011,zyuzin2012,wang2013a,Hosur-anomaly,
Hosur2013,Kim-chiral-anomaly,Parameswaran-anomaly, Zhou-plasmon,Li2015ZrTe5,Goswami2015}. The locations of Weyl points are tunable by the environmental conditions such as pressure and magnetic field, but it is impossible to eliminate a single Weyl point without breaking the translational symmetry of the crystals\cite{zyuzin2012,yang2011,wang2013a,wei2012,Sun2015helical,Bi2015}. The real materials of Weyl semimetal have been discovered very recently\cite{weng2015,Huang2015TaAs,Zhang2015a,xu2015,lv2015,Huang2015,
YangLexian,Ghimire,Shekhar,Xu2015NbAs}(Meanwhile, the photonic analog of Weyl semimetals has also been discovered\cite{lu2015,lu2013weyl,Lu-review}).

Another class of topological semimetal in three spatial dimensions is
the topological nodal line semimetal\cite{Burkov2011nodal,Carter2012,
Phillips2014tunable,chen2015topological,
Zeng2015nodal,Chiu2014,Mullen2015,Weng2015nodal,Yu2015,Kim2015,Bian2015nodal,
Xie2015ring,Rhim2015Landau,Chen2015spin,
Chiu2015classification,Fang2015nodal,Bian2015TlTaSe},
which features a line of band-touching points in the band structure.
Unlike the Weyl points in Weyl semimetals, certain symmetries are
required to ensure the stability of the gapless nodal
lines\cite{Kim2015,Fang2015nodal} in semimetals. Partially gapping
out the nodal lines can leave certain isolated points gapless,
realizing the Weyl semimetal\cite{weng2015}. In this sense, the nodal
line semimetals can be regarded as intermediate systems bridging the
conventional metals and Weyl semimetals. Meanwhile, the nodal line
phases have also been proposed in the photonic
crystals\cite{lu2013weyl,Lu-review}, which pave the way for the
construction of photonic Weyl crystals. Several materials, including
Cu$_3$PdN\cite{Yu2015,Kim2015} and Ca$_3$P$_2$\cite{Xie2015ring},
have emerged as candidates of the nodal line semimetals.

Considering the unique band structures of the nodal line semimetals,
which are significantly different from both the ordinary metals with
large Fermi surfaces and Weyl semimetals with nodal points, it is
natural to study the behavior of collective modes, including the
plasmon. In the case of three-dimensional Dirac and Weyl semimetals,
the plasmon frequency in the long-wavelength limit follows the
$\omega_p\sim n^{1/3}$ law\cite{Sarma2009}, where $\omega_p$ is the plasmon
frequency, and $n$ is the electron density, which is distinct from
the usual $\omega_p\sim n^{1/2}$ behavior of an ordinary electron
liquids. In this paper we study the plasmon in the nodal line
semimetals, and find a $\omega_p\sim n^{1/4}$ law for the plasmon
frequency in the regime that the interband contribution can be neglected. In the regime of very small doping, for which the interband contribution is important, the dispersion crossovers to a $\omega_p\sim n^{1/2}$ law.  The novel features of the collective modes are useful in establishing new candidates for the nodal line semimetals.

The rest of this paper is organized as follows. In Sec.\ref{sec:model} we present our results of the polarizablity. Based on these results we study the behavior of the plasmon in Sec.\ref{sec:plasmon}, obtaining the dependence of plasmon frequency on the electron density.

\section{The model of nodal line semimetal and the polarizability }\label{sec:model}

\begin{figure}
\includegraphics[width=6.0cm, height=4.3cm]{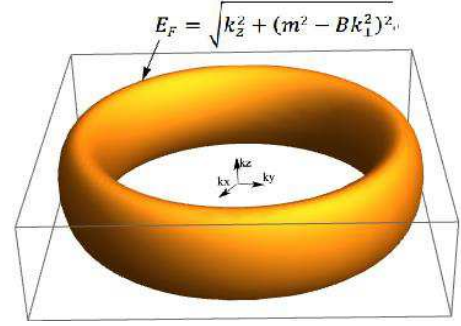}
\caption{ An illustration of the Fermi surfaces of a nodal line
semimetal, for nonzero Fermi energy $E_F$. }  \label{sketch}
\end{figure}

We study a continuous model with the Hamiltonian\cite{Fang2015nodal}
given by $H=\sum_{k}\hat{\Phi}_{k}^{\dag}\mathcal{H}_{0}(k)\hat{\Phi}_{k}$
with $\hat{\Phi}_{k}^{\dag}=(\hat{c}_{k\uparrow}^{\dag},\hat{c}_{k\downarrow}^{\dag})$
and
\begin{eqnarray}
\mathcal{H}_{0}(k)=(m-Bk_{\perp}^{2})\sigma_{x}+k_{z}\sigma_{z},\label{1}
\end{eqnarray}
where $k_{\perp}^{2}=k_{x}^{2}+k_{y}^{2}$, and$\sigma_{x}$ and $\sigma_{z}$
are Pauli matrices, and $m$ and $B$ are positive constants with the dimension of energy and the inverse of energy,
respectively. A sketch of the Fermi surface at nonzero chemical potential is shown in Fig.\ref{sketch}.
This Hamiltonian can be diagonalized by a simple transformation:
\begin{eqnarray}
\left(\begin{array}{c}
        \hat{c}_{k\uparrow} \\
        \hat{c}_{k\downarrow}
      \end{array}
\right)=\left(\begin{array}{cc}
                \cos\theta_{k} & -\sin\theta_{k} \\
                \sin\theta_{k} & \cos\theta_{k}
              \end{array}
\right)\left(\begin{array}{c}
               \hat{c}_{+,k} \\
               \hat{c}_{-,k}
             \end{array}
\right),\label{2}
\end{eqnarray}
with $\theta_{k}=\frac{1}{2}\arctan((m-Bk_{\perp}^{2})/k_{z})$. The diagonalized Hamiltonian reads
\begin{eqnarray}
H=\sum_{k\alpha}\alpha E_{k}\hat{c}_{\alpha,k}^{\dag}\hat{c}_{\alpha,k}\label{3}
\end{eqnarray}
with $E_{k}=\sqrt{(Bk_{\perp}^{2}-m)^{2}+k_{z}^{2}}$ and $\alpha=+,-$.

In the imaginary frequency, the lowest-order polarizability is given as
\begin{eqnarray}
P(\bq,i\omega)&=&-\frac{1}{v}\int_{0}^{\beta}d\tau e^{i\omega\tau}<T_{\tau}\hat{\rho}(\bq,\tau)\hat{\rho}(-\bq,0)>,\label{4}
\end{eqnarray}
where $v$ is the volume of the system, and $\hat{\rho}(q,\tau)=\sum_{k\sigma}\hat{c}_{k+q\sigma}^{\dag}(\tau)\hat{c}_{k\sigma}(\tau)$.
Standard calculations lead to the polarizability at finite temperature ($k_BT\equiv 1/\beta$):
\begin{eqnarray}
P(\bq,\omega)&=&\frac{1}{2v}\sum_{k,\alpha\alpha'}\frac{f(\alpha E_{k+q})-f(\alpha' E_{k})}{\omega+\alpha E_{k+q}-\alpha' E_{k}+i\eta}
(1+\alpha\alpha'\cos2\delta\theta_{kq}),\quad\label{polarizability}
\end{eqnarray}
where $\delta\theta_{kq}= \theta_{k+q}-\theta_{k}$, and $f(E)=1/(1+e^{\beta(E-\mu)})$ is the
Fermi-Dirac distribution function with $\mu$ the chemical potential; $\alpha,\alpha'$ take the
two value $+,-$. A small imaginary part $i\eta$ is introduced to account for the decay of quasi-electrons.  At zero temperature, the chemical potential $\mu$ is equivalent to the Fermi energy $E_{F}$.
In the following, we shall focus on the cases of $E_{F}>0$, while the cases of $E_{F}<0$ are essentially equivalent. The plasmon frequency is determined by \bea \epsilon(\bq,\omega)\equiv 1-v(\bq)P(\bq,\omega)=0, \label{RPA} \eea indicating the presence of self-sustaining collective modes. In this work we shall be most interested in the long wave-length limit
$q\rightarrow0$.

There are four contributions to $P(\bq,\omega)$, corresponding to $\alpha,\alpha'=+,-$ in Eq.\ref{polarizability}. In the zero-temperature limit, only three of them are nonzero, which are the intraband part $P_{++}(q,\omega)$ and the two interband parts $P_{+-}(q,\omega)$
and $P_{-+}(q,\omega)$. In the regime $\max\{2\sqrt{1+\frac{m}{E_{F}}}q_{x,y}, q_{z}\}<\omega<2E_{F}$, we can see that
$\text{Im}P(\bq,\omega)=0$, thus we only need to calculate the real part of $P(\bq,\omega)$.  For the $++$ part $P_{++}(q,\omega)$ we have
\begin{eqnarray}
&& \text{Re}P_{++}(q\rightarrow0,\omega)\nn \\ &=& \text{Re}\left(\frac{1}{2v}\sum_{k}\frac{f(E_{k+q})-f(E_{k})}{\omega+ E_{k+q}- E_{k}+i\eta}
(1+\cos2\delta\theta_{kq})\right)\nonumber\\
&=&\text{Re}\left(\frac{1}{2v}\sum_{k}\frac{\Theta(E_{F}-E_{k+q})-\Theta(E_{F}-E_{k})}{\omega+ E_{k+q}-E_{k}+i\eta}
(1+\cos2\delta\theta_{kq})\right)\nonumber\\
&=&\int_{-m}^{\infty} \frac{dy}{4\pi^{2}}(\frac{2(y+m)y^{2}}{E_{F}^{2}}\frac{q_{\perp}^{2}}{\omega^{2}} \nn \\ &&
+\frac{(E_{F}^{2}-y^{2})}{E_{F}^{2}}\frac{q_{z}^{2}}{B\omega^{2}})
\frac{E_{F}}{\sqrt{E_{F}^{2}-y^{2}}}\Theta(E_{F}^{2}-y^{2}).\label{7}
\end{eqnarray}

When $E_{F}<m$, we obtain that (see the appendix)
\begin{eqnarray}
\text{Re}P_{++}(q\rightarrow0,\omega) = C_{++}^{\perp}q_{\perp}^{2}+C_{++}^{z} q_{z}^{2} \label{8}
\end{eqnarray}
with $q_{\perp}^{2}=q_{x}^{2}+q_{y}^{2}$, \bea C_{++}^{\perp}=\frac{E_{F}^{2}\tilde{m}}{4\pi\omega^{2}}, \label{9}\eea  and
\bea C_{++}^{z}=\frac{E_{F}^{2}}{8\pi\omega^{2}\tilde{B}}, \label{10}\eea where we have defined the dimensionless quantities $\tilde{m}\equiv m/E_{F}$,
and $\tilde{B}\equiv BE_{F}$.

On the other hand, when $E_{F}>m$ we find that
\begin{eqnarray}
{\rm Re} P_{++}(q\rightarrow 0,\omega) &&=\frac{1}{4\pi^{2}}[g_{1}(\tilde{m})\frac{E_{F}^{2}q_{\perp}^{2}}{\omega^{2}}+g_{2}
(\tilde{m})\frac{E_{F}^{2}\tilde{m}q_{\perp}^{2}}{\omega^{2}} \nn \\ && +
g_{3}(\tilde{m})\frac{E_{F}^{2}q_{z}^{2}}{\tilde{B}\omega^{2}}],\label{11}
\end{eqnarray}
where $g_{1,2,3}$ are three dimensionless functions:
\begin{eqnarray}
g_{1}(x)&=&\frac{2}{3}\sqrt{1-x^{2}}(2+x^{2}),\nonumber\\
g_{2}(x)&=&\frac{1}{2}\left(\frac{\pi}{2}+\arcsin x-x\sqrt{1-x^{2}}\right)\nonumber\\
g_{3}(x)&=&\frac{1}{2}\left(\frac{\pi}{2}+\arcsin x+x\sqrt{1-x^{2}}\right)\label{12}
\end{eqnarray}

Another contribution to the polarizability, $P_{+-}$, is given by
\begin{eqnarray}
P_{+-}(q\rightarrow0,\omega)&=&\frac{1}{2v}\sum_{k}\frac{f(E_{k+q})-f(-E_{k})}{\omega+ E_{k+q}+ E_{k}+i\eta}
(1-\cos2\delta\theta_{kq})\nonumber\\
&=&\frac{1}{2v}\sum_{k}\frac{\Theta(E_{F}-E_{k+q})-1}{\omega+ E_{k+q}+E_{k}+i\eta}
(1-\cos2\delta\theta_{kq})\nonumber\\
&=&\frac{1}{2v}\sum_{[k]}\frac{-1}{\omega+ 2E_{k}+i\eta}
(1-\cos2\delta\theta_{kq}),\label{13}
\end{eqnarray}
where $[k]$ denotes the summation (or integral) over $k$ with the variable restricted to the region $E_{k}>E_{F}$.
Similarly, the last contribution to the polarizability is
\begin{eqnarray}
P_{-+}(q\rightarrow0,\omega)=\frac{1}{2v}\sum_{[k]}\frac{1}{\omega-2E_{k}+i\eta}
(1-\cos2\delta\theta_{kq}),\label{14}
\end{eqnarray}

In the $q\rw 0$ limit,  $(1-\cos2\delta\theta_{kq})$ can be
safely expanded as
$(1-\cos2\delta\theta_{kq})=\frac{[(m-Bk^{2})q_{z}
-2Bk_{\perp}q_{\perp}k_{z}\cos\phi]^{2}}{2E_{k}^{4}}\label{expansion}$, where $\phi$ is the angle between the projection of $\bk$ and $\bq$ to the $k_x$-$k_y$ plane, thus we have
\begin{eqnarray}
&&\text{Re}P_{+-}(q\rightarrow0,\omega) \nn \\
&=&-\frac{1}{2v}\sum_{[k]}\frac{\omega+2E_{k}}{(\omega+2E_{k})^{2}+\eta^{2}}
\frac{[(m-Bk_{\perp}^{2})q_{z}-2Bk_{\perp}q_{\perp}k_{z}\cos\phi]^{2}}{2E_{k}^{4}},\qquad\label{15}
\end{eqnarray} and
\begin{eqnarray}
&&\text{Re}P_{-+}(q\rightarrow0,\omega) \nn \\ &=&\frac{1}{2v}\sum_{[k]}\frac{\omega-2E_{k}}{(\omega-2E_{k})^{2}+\eta^{2}}
\frac{[(m-Bk_{\perp}^{2})q_{z}-2Bk_{\perp}q_{\perp}k_{z}\cos\phi]^{2}}{2E_{k}^{4}}.\qquad\label{16}
\end{eqnarray}
The small imaginary part $\eta$ is relevant only for $ P_{-+}$, because there exists a resonance at $\omega=2E_{k}$ in $ P_{-+}$, therefore, we omit $i\eta$ in $\text{Re}P_{++}$ and $\text{Re}P_{+-}$. In addition, we notice that there exists a logarithmic divergence in both $\text{Re}P_{+-}$ and $\text{Re}P_{-+}$, which is due to the presence of the infinite Dirac sea, an artifact of the continuous model, therefore, we  introduce a cutoff as follows:
$k_\perp^2<k_{\perp,c}^{2}$ and $k_z<k_{z,c}$, with $Bk_{\perp,c}^{2}=k_{z,c}=\Lambda$. For the simplicity of notations, we also introduce dimensionless
quantities $\tilde{\omega}=\omega/E_F$, $\tilde{x}\equiv (Bk_{\perp}^{2}-m)/E_{F}$, $\tilde{z}\equiv k_{z}/E_{F}$, $\tilde{\eta}\equiv \eta/E_{F}$ and
$\tilde{\Lambda}=\Lambda/E_{F}$.

In the regime $E_{F}<m$, $\text{Re}P_{+-}$ can be simplified to
\begin{eqnarray}
\text{Re}P_{+-}(q\rightarrow0,\omega)=C_{+-}^{\perp}q_{\perp}^{2}+C_{+-}^{z} q_{z}^{2}
\end{eqnarray}
with
\begin{eqnarray}
C_{+-}^{z}&=&-\frac{1}{16\pi^{2}\tilde{B}}\left[\int_{-1}^{1}d\tilde{x} \int_{\sqrt{1-\tilde{x}^{2}}}^{+\infty}d\tilde{z}+
\left(\int_{1}^{+\infty}+\int_{-\tilde{m}}^{-1}\right)d\tilde{x} \int_{0}^{+\infty}d\tilde{z}\right]\nn \\ && \frac{\tilde{\omega}+2\tilde{E}_{k}}
{(\tilde{\omega}+2\tilde{E}_{k})^{2}}\frac{\tilde{x}^{2}}
{(\tilde{x}^{2}+\tilde{z}^{2})^{2}},
\end{eqnarray}
where $\tilde{E}_{k}=\sqrt{\tilde{x}^{2}+\tilde{z}^{2}}$ (Since the integral is convergent, we have put the upper-limit of integration to infinity), and
\begin{eqnarray}
C_{+-}^{\perp}&=&  \left[\int_{-1}^{1}d\tilde{x}\int_{\sqrt{1-\tilde{x}^{2}}}^{+\infty}d\tilde{z}+
+\int_{-\tilde{m}}^{-1}d\tilde{x}\int_{0}^{+\infty}d\tilde{z} +\int_{1}^{\tilde{\Lambda}}
d\tilde{x}\int_{0}^{\tilde{\Lambda}}d\tilde{z}\right] \nn \\ && \frac{-1}{8\pi^{2}}  \frac{\tilde{\omega}+2\tilde{E}_{k}}
{(\tilde{\omega}+2\tilde{E}_{k})^{2}}\frac{(\tilde{x}+\tilde{m})\tilde{z}^{2}}
{(\tilde{x}^{2}+\tilde{z}^{2})^{2}},
\end{eqnarray}
Similarly, we also find that
\begin{eqnarray}
\text{Re}P_{-+}(q\rightarrow0,\omega)=C_{-+}^{\perp}q_{\perp}^{2}+C_{-+}^{z} q_{z}^{2} \label{20}
\end{eqnarray}
with
\begin{eqnarray}
C_{-+}^{z}&=&\frac{1}{16\pi^{2}\tilde{B}} \left[\int_{-1}^{1}d\tilde{x}\int_{\sqrt{1-\tilde{x}^{2}}}^{+\infty}d\tilde{z}+
\left(\int_{1}^{+\infty}+\int_{-\tilde{m}}^{-1}\right) d\tilde{x}\int_{0}^{+\infty}d\tilde{z}\right]\nn \\ && \frac{\tilde{\omega}-2\tilde{E}_{k}}
{(\tilde{\omega}-2\tilde{E}_{k})^{2}+\tilde{\eta}^{2}}\frac{\tilde{x}^{2}}
{(\tilde{x}^{2}+\tilde{z}^{2})^{2}},\label{21}
\end{eqnarray}
and
\begin{eqnarray}
C_{-+}^{\perp}&=&\left[\int_{-1}^{1}d\tilde{x}\int_{\sqrt{1-\tilde{x}^{2}}}^{+\infty}d\tilde{z}+
\int_{-\tilde{m}}^{-1}d\tilde{x}\int_{0}^{+\infty}d\tilde{z}+\int_{1}^{\tilde{\Lambda}}
d\tilde{x}\int_{0}^{\tilde{\Lambda}}d\tilde{z}\right]\nn\\ &&\frac{1}{8\pi^{2}}\frac{\tilde{\omega}-2\tilde{E}_{k}}
{(\tilde{\omega}-2\tilde{E}_{k})^{2}+\tilde{\eta}^{2}}\frac{(\tilde{x}+\tilde{m})\tilde{z}^{2}}
{(\tilde{x}^{2}+\tilde{z}^{2})^{2}},\label{22}
\end{eqnarray}

To suppress the formalism, let us define
\begin{eqnarray}
C_{T}^{z} = C_{+-}^{z}+C_{-+}^{z},\nonumber\\
C_{T}^{\perp} = C_{+-}^{\perp}+C_{-+}^{\perp}.
\end{eqnarray}
Now we can made an expansion in $\tilde{\omega}$ and
obtain the following analytic results:
\begin{eqnarray}
C_{T}^{z}  =-\frac{1}{16\pi^{2}\tilde{B}}\left[(\frac{\pi}{2}-\frac{2}{3\tilde{m}})
+(\frac{\pi}{24}-\frac{2}{45\tilde{m}^{3}})\tilde{\omega}^{2}\right],\nonumber\\
C_{T}^{\perp}  =-\frac{1}{8\pi^{2}}\left[\left(C_{1}(\tilde{m})+f(\tilde{\Lambda})\right)
+\left(\frac{\pi}{24}+\frac{1}{180\tilde{m}^{2}}\right)\tilde{\omega}^{2}\right],\label{23}
\end{eqnarray}
where $C_{1}(\tilde{m})=\frac{\pi\tilde{m}}{2}-\frac{\log \tilde{m}}{3}+C$,
with $C=\frac{1}{3\sqrt{2}}-\frac{1}{3}-\text{arc}\sinh1=-0.98$, and $f(\tilde{\Lambda})=\text{arc}\sinh\tilde{\Lambda} -\frac{\tilde{\Lambda}}{\sqrt{1+\tilde{\Lambda}^{2}}}$. We remark that $f(\tilde{\Lambda})$
is a monotonically increasing function, and $f(\tilde{\Lambda})\sim \log\tilde{\Lambda}$ when $\tilde{\Lambda}$ is large\footnote{The logarithmical growth is quite slow, $e.g.$, $f(100)=4.3$, and $f(1000)=6.6$}.

\begin{figure}
\subfigure{\includegraphics[width=7.0cm, height=5cm]{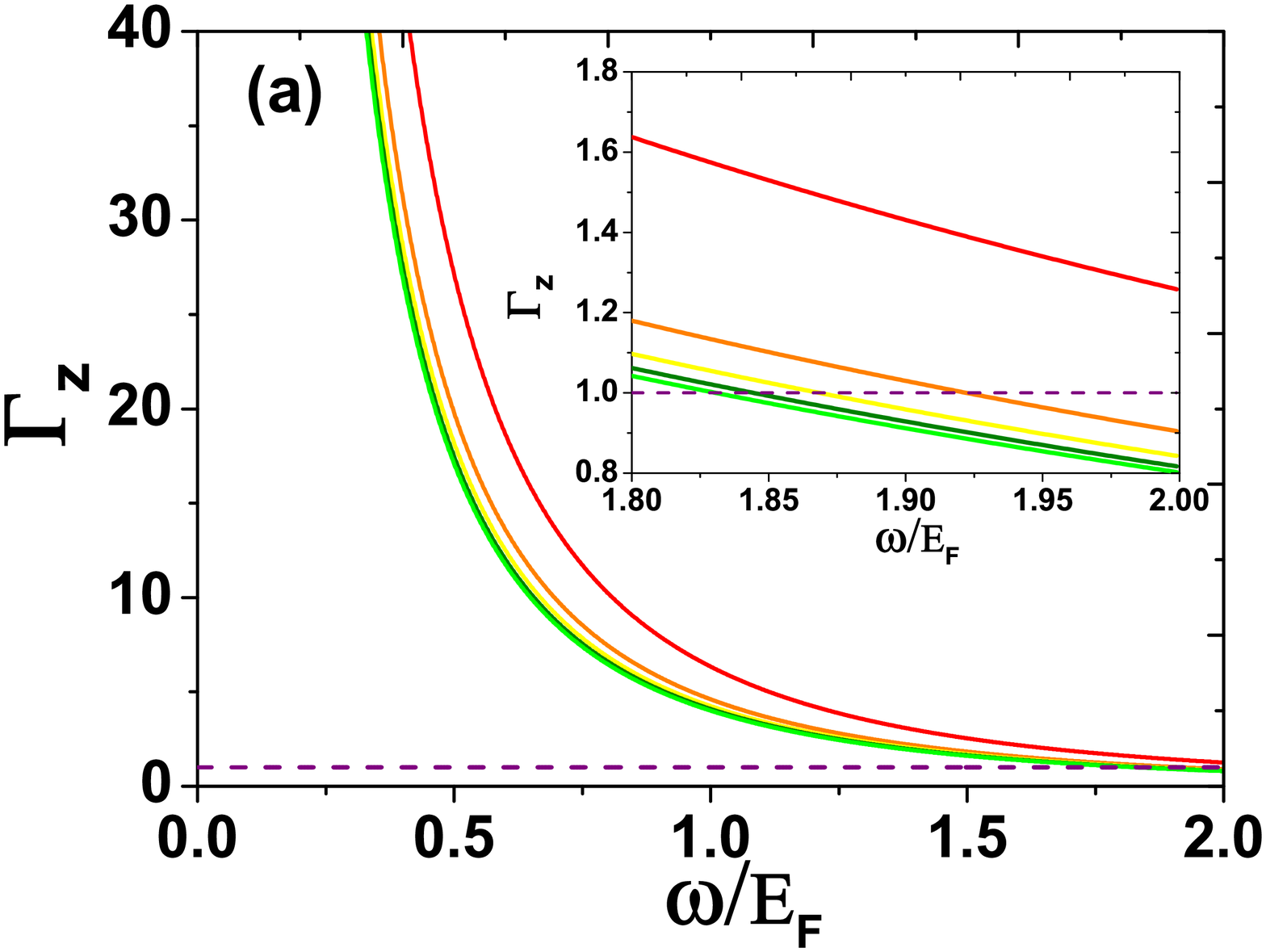}}
\subfigure{\includegraphics[width=7.0cm, height=5cm]{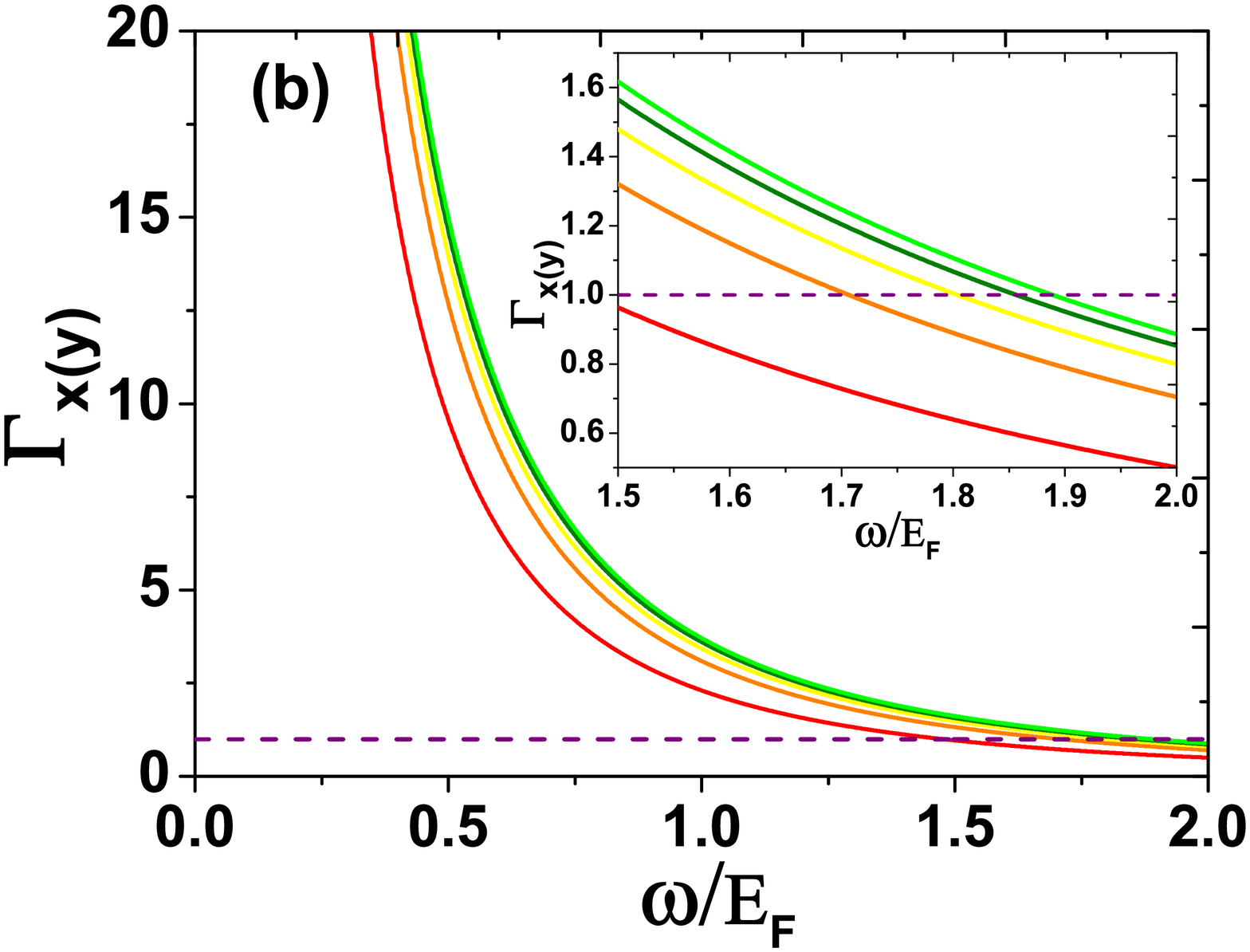}}
\caption{ (a) $\Gamma_{z}$-\textit{vs}-$\tilde{\omega}$. From red line
to blue line, $\tilde{m}=1,2,3,4,5$.  (b) $\Gamma_{\perp}$-\textit{vs}-$\tilde{\omega}$,
the color labeling scheme is the same as (a). For both (a) and (b), $\tilde{\Lambda}=10$,
and the dashed line corresponds to $\Gamma_{z(\perp)}=1$. } \label{fig1}
\end{figure}

Now that the polarizability $P(\bq,\omega)$ have both the intraband part and the interband part, we would like to quantify their ratio by
\begin{eqnarray}
\Gamma_{z}(\tilde{\omega})&=&|\frac{C_{++}^{z}}{C_{T}^{z}}|,\nonumber\\
\Gamma_{\perp}(\tilde{\omega})&=&|\frac{C_{++}^{\perp}}{C_{T}^{\perp}}|.\label{24}
\end{eqnarray} As long as $\Gamma_z$ and $\Gamma_\perp>>1$, the intraband contribution $P_{++}$ dominates.
As shown in Fig.(\ref{fig1}), when $\tilde{\omega}<<2$,  both $\Gamma_{z}$ and $\Gamma_{\perp}>>1$, which indicates that the intraband process dominates the polarizability. On the other hand, if $\tilde{\omega}\rw 2$, then the interband contribution is comparable to the intraband process. It is interesting to
note that the value of $\tilde{m}$ have opposite effect in the $z$-direction and the in-plane directions. As we increase $\tilde{m}$,
the region of $\tilde{\omega}$ for which $\Gamma_{z}<1$ expands, while the region for which $\Gamma_{\perp}<1$ shrinks.

Because the interband contributions has opposite sign to the intraband contribution, i.e. they are negative,  $\Gamma_{z}<1$
or $\Gamma_{\perp}<1$ indicates the absence of plasmonic modes in the respective directions (see the next section).

\section{Plasmon frequency}\label{sec:plasmon}

Based on the results above, the total polarization function $P(q\rightarrow0,\omega)$ in the region
$0<\tilde{\omega}<2$ is given as
\begin{eqnarray}
P(q\rightarrow0,\omega)&=&(1-\Gamma_{\perp}^{-1})C^{\perp}_{++}q_{\perp}^{2} +(1-\Gamma_{z}^{-1})C^{z}_{++}q_{z}^{2}\nonumber\\
&=&\frac{(1-\Gamma_{\perp}^{-1})} {4\pi}\frac{E_{F}m}{\omega^{2}}q_{\perp}^{2} +\frac{(1-\Gamma_{z}^{-1})}{8\pi}\frac{E_{F}}{B\omega^{2}}q_{z}^{2}.\label{25}
\end{eqnarray}
Inserting this equation into the RPA equation, Eq.(\ref{RPA}), with $v(q)=\frac{4\pi e^{2}}{\kappa q^{2}}$,
we obtain the plasmon frequency
\begin{eqnarray}
\omega_{p,\perp}&=&\sqrt{\frac{(1-\Gamma_{\perp}^{-1})e^{2}m}{\kappa}}\sqrt{E_{F}},\nonumber\\
\omega_{p,z}&=&\sqrt{\frac{(1-\Gamma_{z}^{-1})e^{2}}{2B\kappa}}\sqrt{E_{F}}.\label{26}
\end{eqnarray}
Defining the fine structure constant $\alpha_{s}=e^{2}/\kappa$, we have $\omega_{p,\perp}=\sqrt{(1-\Gamma_{\perp}^{-1})\alpha_{s}m}\sqrt{E_{F}}$ and $\omega_{p,z}=\sqrt{\frac{(1-\Gamma_{z}^{-1})\alpha_{s}}{2B}}\sqrt{E_{F}}$,
which are in fact a self-consistent equation because $\Gamma_{\perp}$ and
$\Gamma_{z}$ themselves are functions of $\omega$ and $E_{F}$.
More explicitly, and in terms of the dimensionless quantities,
\begin{eqnarray}
\tilde{\omega}_{p,\perp}&=& \sqrt{(1-\Gamma_{\perp}^{-1}(\tilde{\omega}_{p,\perp}))\alpha_{s}\tilde{m}},\nonumber\\
\tilde{\omega}_{p,z}&=& \sqrt{\frac{(1-\Gamma_{z}^{-1}(\tilde{\omega}_{p,z}))\alpha_{s}}{2\tilde{B}}},\label{self-c}
\end{eqnarray}

\begin{figure}
\subfigure{\includegraphics[width=7.0cm, height=4.5cm]{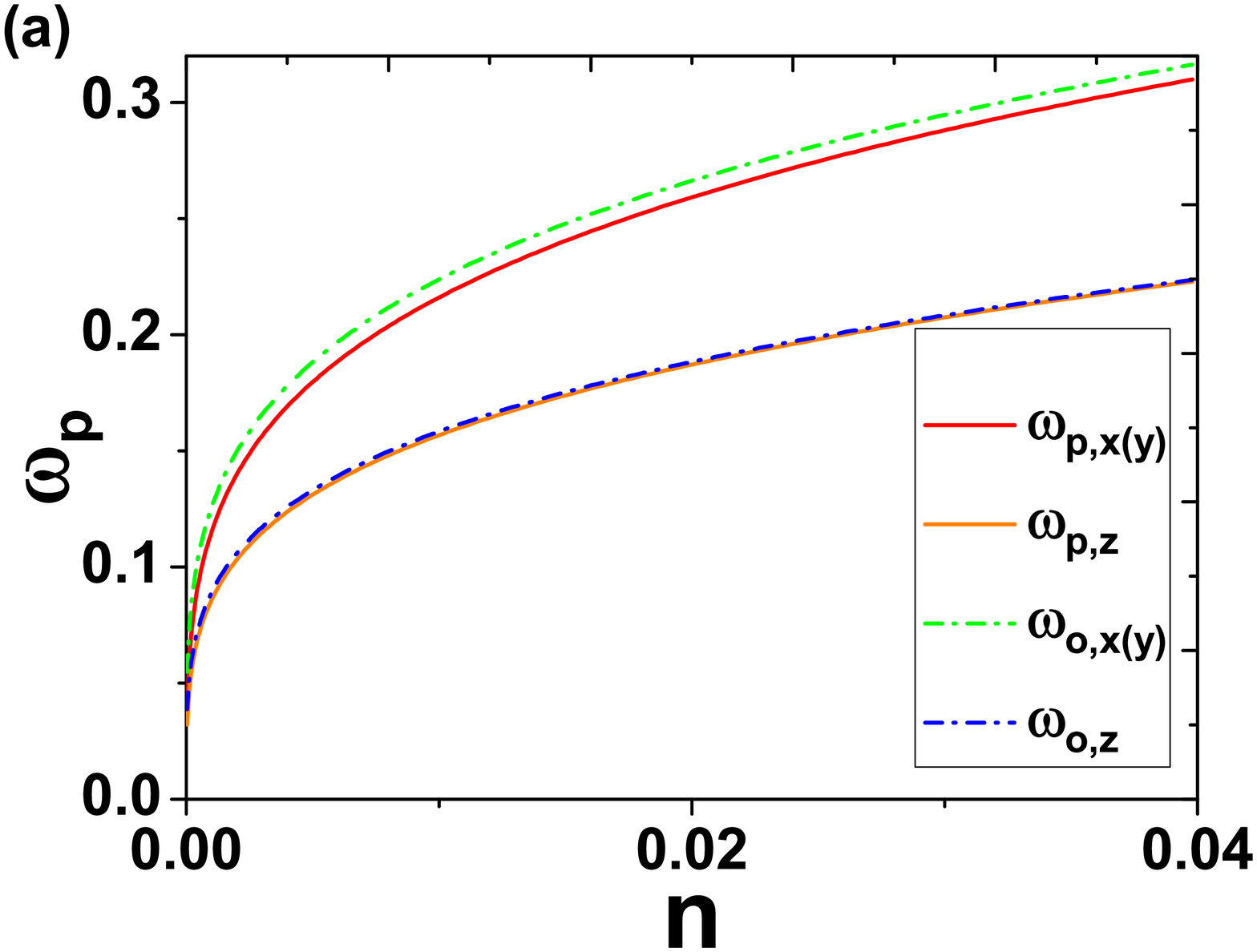}}
\subfigure{\includegraphics[width=7.0cm, height=4.5cm]{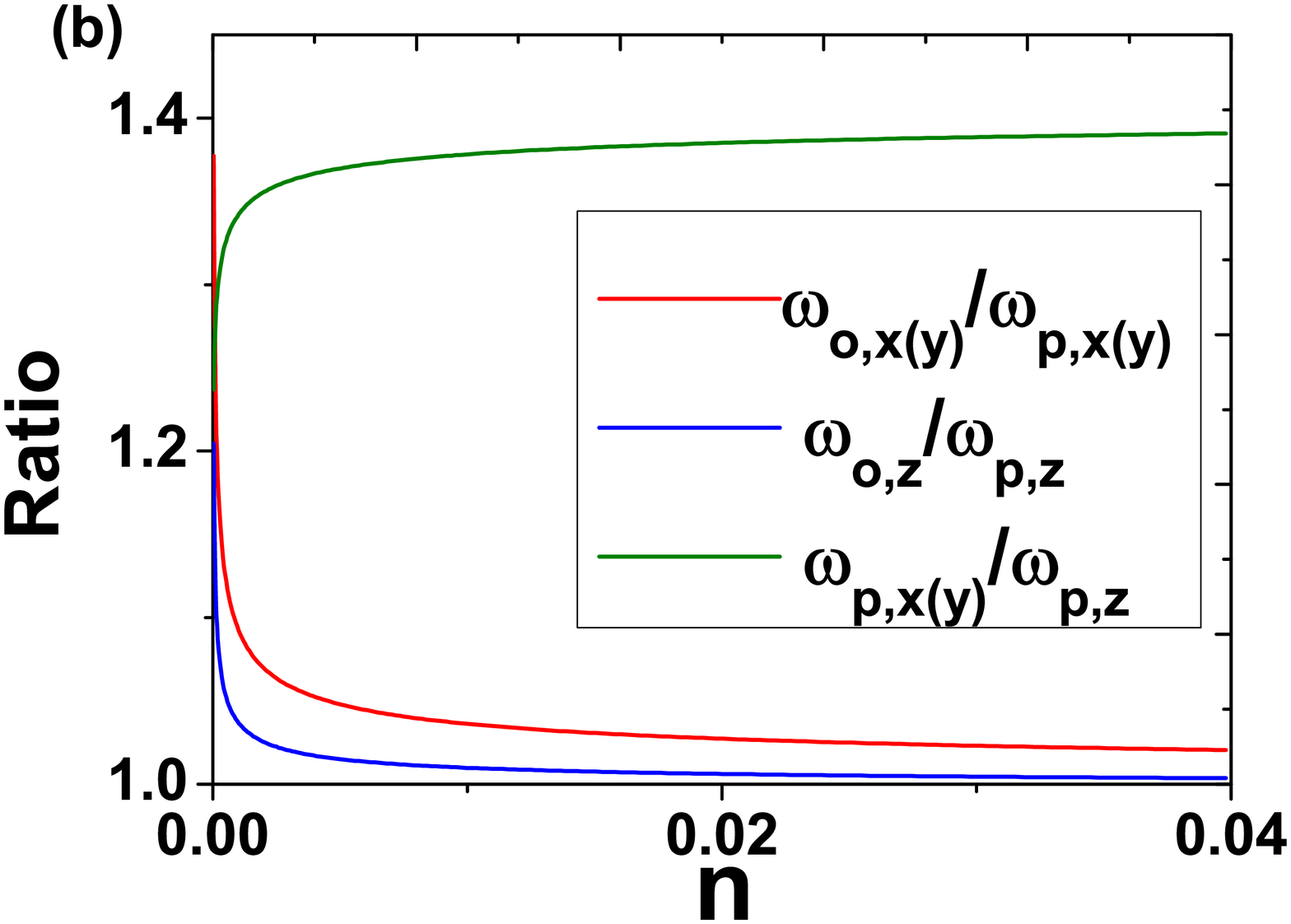}}
\subfigure{\includegraphics[width=7.0cm, height=4.5cm]{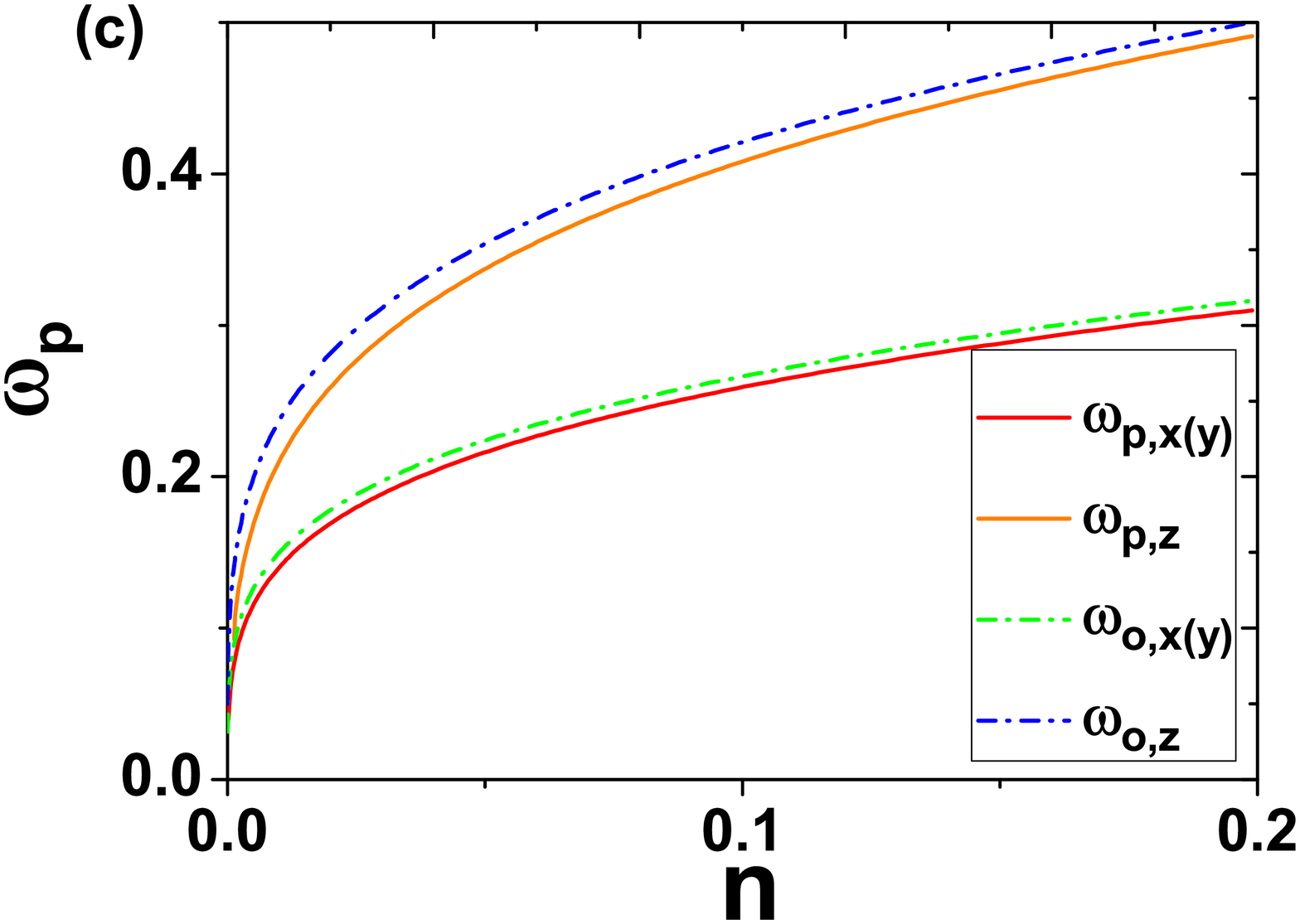}}
\subfigure{\includegraphics[width=7.0cm, height=4.5cm]{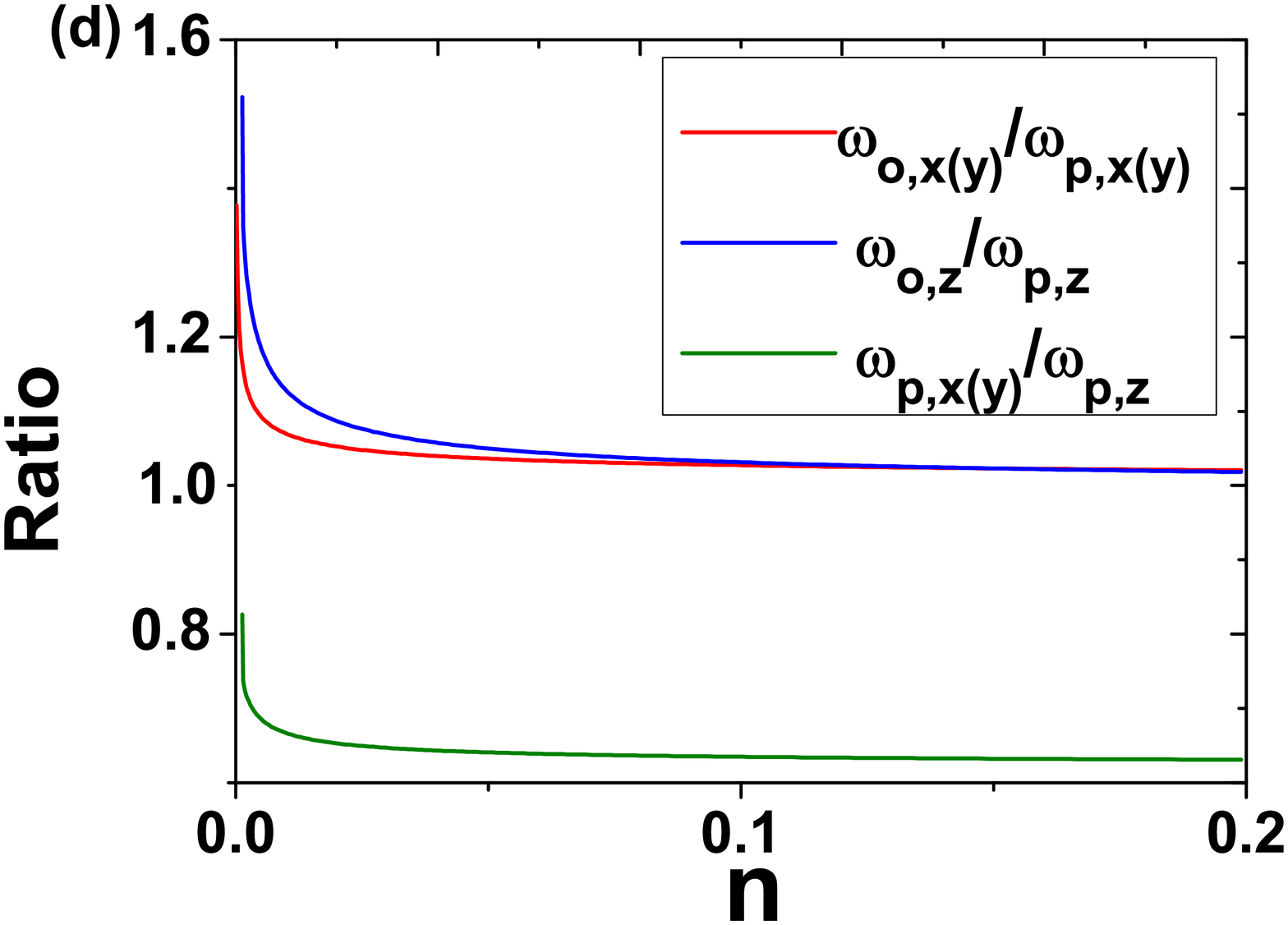}}
\caption{ The $\omega$-vs-$n$ plot and the ratios among the relevant frequencies, for two values of $B$: $B=1$ for (a)(b), and $B=0.2$ for (c)(d). Common parameters: $\alpha_s=0.1$, $m=1$, $\Lambda=10m$, $E_{F}$ is varied from 0.03 to 1. Here $\omega_{p,x(y)}$ and $\omega_{p,z}$ denote the plasmon frequencies calculated self-consistently (see the text), while $\omega_{o,x(y)}$ and
$ \omega_{o,z} $ denote the plasmon frequencies determined by taking into account only
the intraband part of the polarizability, $i.e.$, $\Gamma_{\perp(z)}^{-1}$ is
set to zero.  From (b) and (d), it is clear that $\omega_o/\omega_p\approx 1$ as long as $n$ is not very small, therefore, it is accurate to omit the interband contribution for most of the values of $n$. } \label{fig2}
\end{figure}

We shall focus on the $E_F<m$ regime, which is the most interesting  regime, since we are most concerned with the physics dominated by the nodal line (When $E_F>m$, the topology of the Fermi surface is no longer a torus).
When $E_{F}<m$, the electron density can be obtained as  \bea n=\frac{E_{F}^{2}}{8\pi B},  \eea  therefore,
we obtain
\begin{eqnarray}
\omega_{p,\perp}&=&\left(\sqrt{\sqrt{8\pi B}(1-\Gamma_{\perp}^{-1})\alpha_{s}m}\right)n^{1/4},\nonumber\\
\omega_{p,z}&=&\left(\sqrt{\sqrt{\frac{2\pi}{B}}(1-\Gamma_{z}^{-1})\alpha_{s}}\right)n^{1/4}.\label{28}
\end{eqnarray} The dependence of $\omega_{p,\perp}$ and $\omega_{p,z}$ for two sets of parameters are shown in Fig.\ref{fig2}.

It is readily seen that
\begin{eqnarray}
\frac{\omega_{p,\perp}}{\omega_{p,z}}=\sqrt{2Bm}\sqrt{\frac{(1-\Gamma_{\perp}^{-1})}{(1-\Gamma_{z}^{-1})}}.\label{29}
\end{eqnarray}

\begin{figure}
\subfigure{\includegraphics[width=7.0cm, height=5cm]{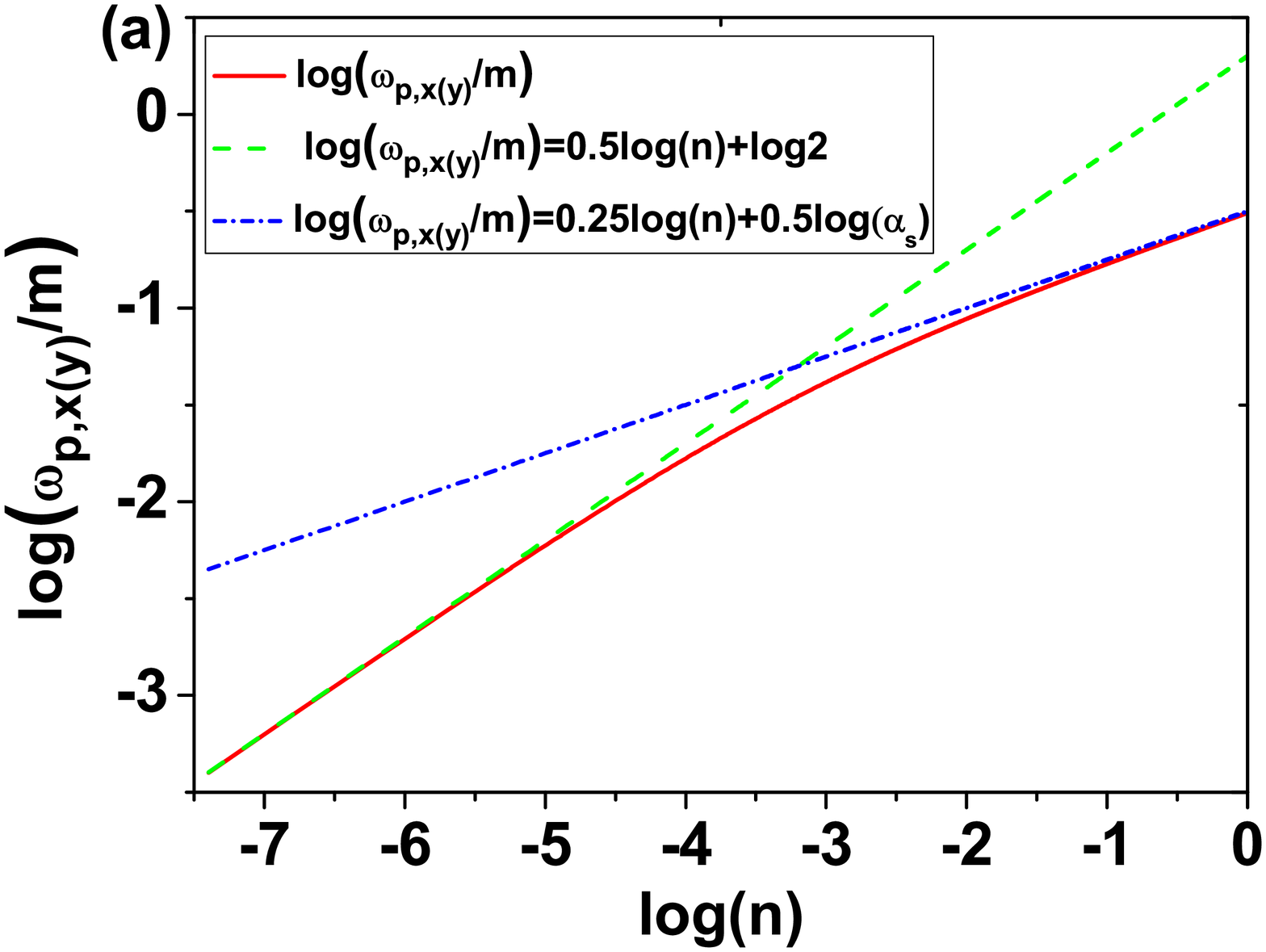}}
\subfigure{\includegraphics[width=7.0cm, height=5cm]{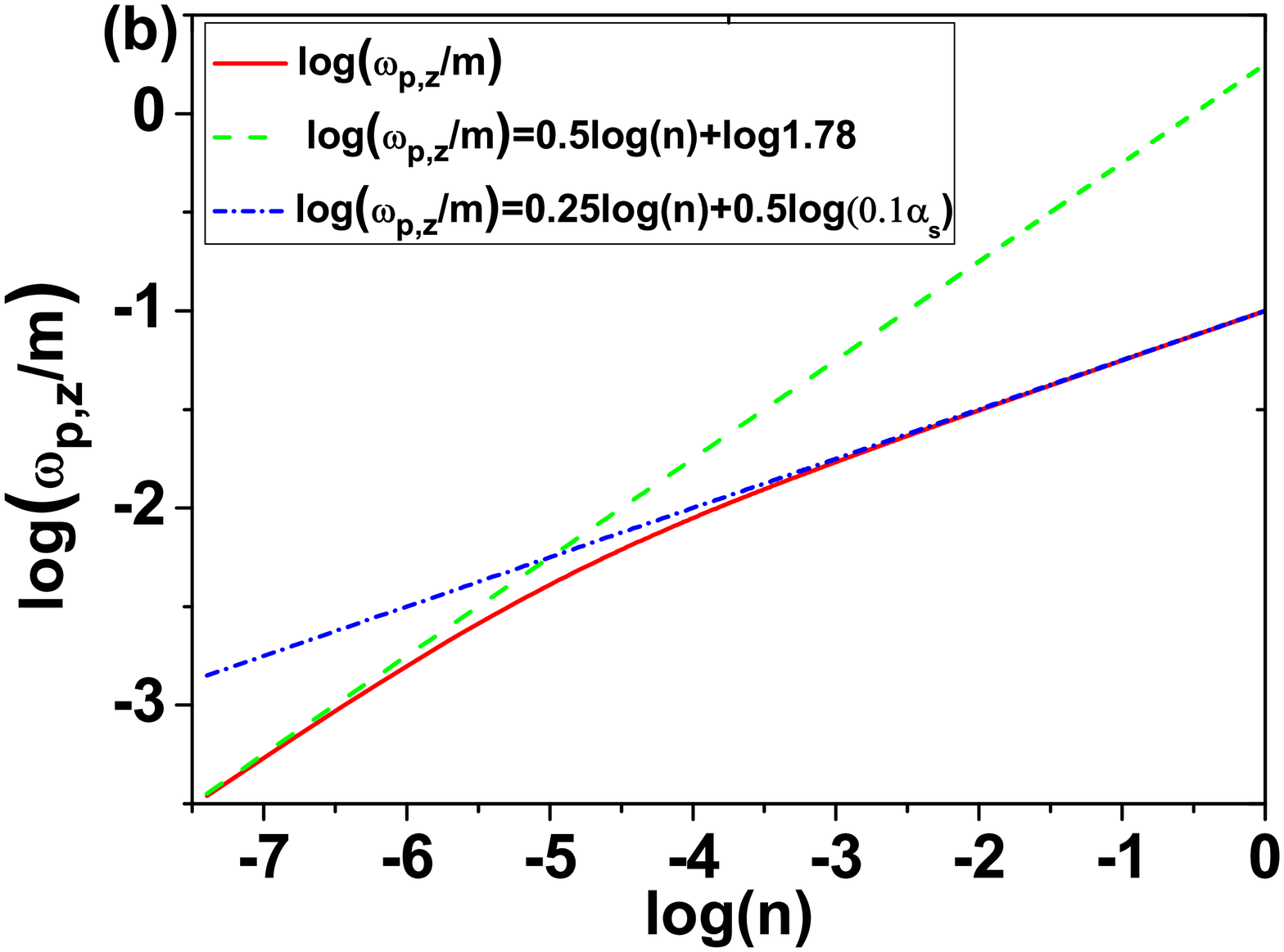}}
\caption{ The $\log\omega_{p}$-$\log n$ plot. Common parameters:
$\alpha=0.1$, $mB=5$, $m^{2}=8\pi B$, $\Lambda/m=10$, and $n=(E_{F}/m)^{2}$. The red solid lines are the self-consistent solution of the plasmon frequencies.
The green dashed lines are the fitting to Eq.(\ref{34}) (the $n^{1/2}$ power-law), while the
blue dash-dot lines are the fitting to Eq.(\ref{32}) (the $n^{1/4}$ power-law). We can see that the plasmon frequencies crossovers from the $n^{1/2}$ law in the regime of very small doping to the $n^{1/4}$ law in the regime of larger doping.  } \label{fig3}
\end{figure}

If the contribution from  interband process can be neglected, $i.e.$, $\Gamma_{\perp(z)}^{-1}$ is taken as zero,
then it is found that
\begin{eqnarray}
\omega_{p,\perp}&=&\left(\sqrt{\sqrt{8\pi B}\alpha_{s}m}\right)n^{1/4},\nonumber\\
\omega_{p,z}&=&\left(\sqrt{\sqrt{\frac{2\pi}{B}}\alpha_{s}}\right)n^{1/4},\label{32}
\end{eqnarray}
The $\omega_p\sim n^{1/4}$ law of plasmon frequency is among the central results of this paper.
As shown in Fig.\ref{fig2} and Fig.\ref{fig3}, the self-consistently determined plasmon frequency agrees with the
$\omega_p\sim n^{1/4}$ law as long as the doping level is not very low.
It is well known that the plasmon frequency follows the $n^{1/2}$ law in an ordinary electron liquid,
and it has also been found that the plasmon frequency of a massless Dirac electron liquid follows
the $n^{1/3}$ law\cite{Sarma2009}. In contrast to these materials, we find a $\omega_p\sim n^{1/4}$
law for the plasmon in the nodal line semimetals.

In very low doping regime, $i.e.$, $n$, or say $E_{F}$, is quite small, from Fig.\ref{fig2}(b)(d)
we can already see that the interband effect becomes important, consequently, the
power-law relation will deviate from the $\omega_p\sim n^{1/4}$ law which holds in most doping regime.
To see this more explicitly, we consider the very low doping regime where $\tilde{m}>>1$, $\tilde{B}<<1$
and derive the analytic expressions of the plasmon frequencies.
As $\tilde{m}>>1$, $\tilde{B}<<1$,  from
Eq.(\ref{28}), it is readily seen that the condition for existence of self-consistent
solution in the regime $0<\tilde{\omega}<2$ puts the following constraints on
$\Gamma_{p,\perp}$ and $\Gamma_{p,z}$:
\begin{eqnarray}
&&1-\frac{4}{\alpha_{s}\tilde{m}}<\Gamma_{p,\perp}^{-1}<1,\nonumber\\
&&1-\frac{8\tilde{B}}{\alpha_{s}}<\Gamma_{p,z}^{-1}<1.\label{33}
\end{eqnarray}
In the very low doping regime, the two constraints indicates that
both $\Gamma_{p,\perp}$ and $\Gamma_{p,z}$ goes to 1. With the
assumption that $\Lambda/m$ is fixed and $\tilde{m}>>1/\tilde{B}$, a combination of Eq.(\ref{9}),
Eq(\ref{10}), Eq.(\ref{23}), Eq.(\ref{24}) and Eq.(\ref{self-c}) gives
the self-consistent solutions, to the first order of the two small
quantities $1/\alpha_{s}\tilde{m}$ and $\tilde{B}/\alpha_{s}$
(details are given in Appendix C):
\begin{eqnarray}
&&\tilde{\omega}_{p,\perp}=\tilde{\omega}_{\perp,c}-\frac{\tilde{\omega}_{\perp,c}^{3}}{2}\frac{1}{\alpha_{s}\tilde{m}},\nonumber\\
&&\tilde{\omega}_{p,z}=\tilde{\omega}_{z,c}-\frac{24\tilde{\omega}_{z,c}}{7}\frac{\tilde{B}}{\alpha_{s}}\label{34}
\end{eqnarray}
with $\tilde{\omega}_{\perp,c}=2/\sqrt{1+(2\log(\tilde{\Lambda})/\pi\tilde{m})}\approx 2$
and $\tilde{\omega}_{z,c}=\sqrt{2(\sqrt{21}-3)}\approx1.78$. With the quite small first order term neglected,
Eqs.(\ref{34}) indicate that in the very low doping regime,
\bea \omega_{p,\perp(z)}\sim E_{F}\sim n^{1/2}. \eea To be more quantitative, we
perform numerical calculations of the original Eq.(\ref{self-c}) for a broad range of of $n$,
and fit the plasmon frequencies to Eq.(\ref{32}) and Eq.(\ref{34})
[see Fig.\ref{fig3}]. In the very low density regime,
$\log(\omega_{p,\perp})$ and $\log(\omega_{p,z})$ are well
fitted by the green dashed lines whose slope is $1/2$, indicating $\omega_{p,\perp(z)}\sim n^{1/2}$;
while in the regime $\log(n) \gtrsim -2$, $i.e.$, $E_{F}/m\gtrsim 0.1$, both $\log(\omega_{p,\perp})$
and $\log(\omega_{p,z})$  are well fitted by the blue dash-dot lines whose slope is $1/4$,
indicating $\omega_{p,\perp(z)}\sim n^{1/4}$. In other words, the plasmon frequencies crossovers from the $n^{1/2}$ law in the regime of very small doping, for which the interband contribution is important, to the $n^{1/4}$ law in the regime of larger doping, for which the intraband contribution dominates.


Finally, we remark that the plasmon frequency is also proportional to $n^{1/4}$ in graphene\cite{Hwang2007}, however, there is a major difference between the plasmon in the graphene and that in the nodal line semimetal. For the graphene, the plasmon is a gapless mode in the long wavelength limit (Its frequency is proportional to $n^{1/4}q^{1/2}$); for the latter, the plasmon is gapped. The difference has a simple origin. Although the Coulomb potential takes the same real-space form of $1/r$ in both the case of graphene in two dimensions and the case of nodal line semimemetal in three dimensions, it is quite different in the momentum space ($q$-space): in three dimensions the $q$-space Coulomb potential is $v_{3d}(q)\propto q^{-2}$, while in two-dimensions it is $v_{2d}(q)\propto q^{-1}$. The plasmon frequencies determined by Eq.(\ref{RPA}) is thus significantly different in two-dimensions and three-dimensions, in particular, the two-dimensional plasmon is gapless, while the three-dimensional plasmon is gapped.

\section{Conclusions}

In conclusion, we have studied the plasmon modes in the nodal line semimetals. The dependence of plasmon frequency on the electron density generally follows a $\omega_p\sim n^{1/4}$ law in the regime of modest doping, in contrast to the ordinary electron liquids and the Dirac/Weyl electron liquids. In the regime of very small doping, for which the interband contribution is important, this $\omega_p\sim n^{1/4}$ law crossovers to a $\omega_p\sim n^{1/2}$ law. The unique features of plasmons will be useful in identifying topological nodal line semimetals using the plasmonics methods.

\emph{Acknowledgements.} This work is supported by NSFC under Grant
No. 11304175 and the Tsinghua University Initiative Scientific
Research Program.

Note added: Upon finishing this paper, we became aware of a closely related
work\cite{Rhim2015plasmon} by Jun-Won Rhim and Yong Baek Kim.

\appendix

\section{Calculations of the polarizability}

The interband contribution $\text{Re}P_{++}(q,\omega)$ can be obtained as follows:

\begin{eqnarray}
&&\text{Re}P_{++}(q\rightarrow0,\omega)\nn \\
&=&\text{Re}\left(\frac{1}{2v}\sum_{k}\frac{f(E_{k+q})-f(E_{k})}{\omega+ E_{k+q}- E_{k}+i\eta}
(1+\cos2\delta\theta_{kq})\right)\nonumber\\
&=&\text{Re}\left(\frac{1}{2v}\sum_{k}\frac{\Theta(E_{F}-E_{k+q})-\Theta(E_{F}-E_{k})}{\omega+ E_{k+q}-E_{k}+i\eta}
(1+\cos2\delta\theta_{kq})\right)\nonumber\\
&\simeq&-\frac{1}{v}\sum_{k}\frac{\delta(E_{F}-E_{k})
\frac{\partial E_{k}}{\partial k_{\beta}}q_{\beta}}{\omega+\frac{\partial E_{k}}{\partial k_{\gamma}}q_{\gamma}}
\nonumber\\
&\simeq&-\frac{1}{v}\sum_{k}(1-\frac{\partial E_{k}}{\partial k_{\beta}}\frac{q_{\beta}}{\omega})
\frac{\partial E_{k}}{\partial k_{\gamma}}\frac{q_{\gamma}}{\omega}\delta(E_{F}-E_{k})\nonumber\\
&=&\int \frac{d^{3}k}{(2\pi)^{3}}(4[\frac{B^{2}k_{x}^{2}}{E_{F}^{2}}\frac{q_{x}^{2}}{\omega^{2}}
+\frac{B^{2}k_{y}^{2}}{E_{F}^{2}}\frac{q_{y}^{2}}{\omega^{2}}](m-Bk_{\perp}^{2})^{2}+
\frac{k_{z}^{2}}{E_{F}^{2}}\frac{q_{z}^{2}}{\omega^{2}})
\frac{E_{F}}{k_{z,0}}\nonumber\\
&&\quad\times(\delta(k_{z}-k_{z,0})+\delta(k_{z}+k_{z,0}))\Theta(E_{F}^{2}-(m-Bk_{\perp}^{2})^{2})\nonumber\\
&=&\int_{0}^{\infty} \frac{dk_{\perp}^{2}}{4\pi^{2}}(\frac{2B^{2}k_{\perp}^{2}}{E_{F}^{2}}\frac{q_{\perp}^{2}}{\omega^{2}}(m-Bk_{\perp}^{2})^{2}
+\frac{(E_{F}^{2}-(m-Bk_{\perp}^{2})^{2})}{E_{F}^{2}}\frac{q_{z}^{2}}{\omega^{2}})\nonumber\\
&&\quad\times\frac{E_{F}}{\sqrt{E_{F}^{2}-(m-Bk_{\perp}^{2})^{2}}}\Theta(E_{F}^{2}-(m-Bk_{\perp}^{2})^{2})\nonumber\\
&=&\int_{0}^{\infty} \frac{dx}{4\pi^{2}}(\frac{2x(x-m)^{2}}{E_{F}^{2}}\frac{q_{\perp}^{2}}{\omega^{2}}
+\frac{(E_{F}^{2}-(m-x)^{2})}{E_{F}^{2}}\frac{q_{z}^{2}}{B\omega^{2}})
\nonumber\\
&&\quad\times\frac{E_{F}}{\sqrt{E_{F}^{2}-(m-x)^{2}}}\Theta(E_{F}^{2}-(m-x)^{2})\nonumber\\
&=&\int_{-m}^{\infty} \frac{dy}{4\pi^{2}}(\frac{2(y+m)y^{2}}{E_{F}^{2}}\frac{q_{\perp}^{2}}{\omega^{2}}
+\frac{(E_{F}^{2}-y^{2})}{E_{F}^{2}}\frac{q_{z}^{2}}{B\omega^{2}})\nonumber\\
&&\quad\times\frac{E_{F}}{\sqrt{E_{F}^{2}-y^{2}}}\Theta(E_{F}^{2}-y^{2})
\end{eqnarray}
For the simplicity of notation, we have defined  $k_{z,0}=\sqrt{E_{F}^{2}-(m-Bk_{\perp}^{2})^{2}}$, $x=Bk_{\perp}^{2}$, and $y=x-m$.
When two subscripts are same, the Einstein summation convention is assumed.

When $E_{F}<m$, we obtain
\begin{eqnarray}
&&\text{Re}P_{++}(q\rightarrow0,\omega)\nn \\ &=&\int_{-E_{F}}^{+E_{F}} \frac{dy}{4\pi^{2}}(\frac{4(y+m)y^{2}}{2E_{F}^{2}}\frac{q_{\perp}^{2}}{\omega^{2}}
+\frac{(E_{F}^{2}-y^{2})}{E_{F}^{2}}\frac{q_{z}^{2}}{B\omega^{2}})
\frac{E_{F}}{\sqrt{E_{F}^{2}-y^{2}}}\nonumber\\
&=&\int_{-\pi/2}^{+\pi/2} \frac{d\theta}{4\pi^{2}}E_{F}(2(E_{F}\sin\theta+m)\sin^{2}\theta
\frac{q_{\perp}^{2}}{\omega^{2}}
+\cos^{2}\theta\frac{q_{z}^{2}}{B\omega^{2}})\nonumber\\
&=&\frac{1}{8\pi}(\frac{2E_{F}mq_{\perp}^{2}}{\omega^{2}}+\frac{E_{F}q_{z}^{2}}{B\omega^{2}})\nonumber\\
&=&C_{++}^{\perp}q_{\perp}^{2}+C_{++}^{z} q_{z}^{2}
\end{eqnarray}
with $C_{++}^{\perp}=\frac{E_{F}^{2}\tilde{m}}{4\pi\omega^{2}}$ and
$C_{++}^{z}=\frac{E_{F}^{2}}{8\pi\omega^{2}\tilde{B}}$.

When $E_{F}>m$, we obtain
\begin{eqnarray}
&& \Re P_{++}(q\rightarrow0,\omega)\nn\\ &=&\int_{-m}^{+E_{F}} \frac{dy}{4\pi^{2}}(\frac{4(y+m)y^{2}}{2E_{F}^{2}}\frac{q_{\perp}^{2}}{\omega^{2}}
+\frac{(E_{F}^{2}-y^{2})}{E_{F}^{2}}\frac{q_{z}^{2}}{B\omega^{2}})
\frac{E_{F}}{\sqrt{E_{F}^{2}-y^{2}}}\nonumber\\
&=&\int_{-\arcsin\frac{m}{E_{F}}}^{+\pi/2} \frac{d\theta}{4\pi^{2}}(2(E_{F}\sin\theta+m)E_{F}\sin^{2}\theta
\frac{q_{\perp}^{2}}{\omega^{2}}
+E_{F}\cos^{2}\theta\frac{q_{z}^{2}}{B\omega^{2}})
\nonumber\\
&=&\frac{1}{4\pi^{2}}(g_{1}(\frac{m}{E_{F}})\frac{E_{F}^{2}q_{\perp}^{2}}{\omega^{2}}+g_{2}
(\frac{m}{E_{F}})\frac{E_{F}mq_{\perp}^{2}}{\omega^{2}}+
g_{3}(\frac{m}{E_{F}})\frac{E_{F}q_{z}^{2}}{B\omega^{2}}),
\end{eqnarray}
where $g_{1,2,3}$ are three dimensionless functions with
\begin{eqnarray}
g_{1}(x)&=&\frac{2}{3}\sqrt{1-x^{2}}(2+x^{2}),\nonumber\\
g_{2}(x)&=&\frac{1}{2}\left(\frac{\pi}{2}+\arcsin x-x\sqrt{1-x^{2}}\right)\nonumber\\
g_{3}(x)&=&\frac{1}{2}\left(\frac{\pi}{2}+\arcsin x+x\sqrt{1-x^{2}}\right)
\end{eqnarray}

\section{Electron density}

The density of conduction-band electron is given by
\begin{eqnarray}
n
&=&\int \frac{d^{3}k}{(2\pi)^{3}}\theta(E_{F}-E_{k})\nonumber\\
&=& \frac{1}{(2\pi)^{2}}\int_{0}^{+\infty}k_{\perp}dk_{\perp}\int_{-\sqrt{E_{F}^{2}-(m-Bk_{\perp}^{2})^{2}}}^{\sqrt{E_{F}^{2}-(m-Bk_{\perp}^{2})^{2}}}dk_{z}
\theta(E_{F}-|m-Bk_{\perp}^{2}|)\nonumber\\
&=&\frac{1}{(2\pi)^{2}}\int_{0}^{+\infty}dk_{\perp}^{2}\sqrt{E_{F}^{2}-(m-Bk_{\perp}^{2})^{2}}\theta(E_{F}-|m-Bk_{\perp}^{2}|)\nonumber\\
&=&\frac{1}{(2\pi)^{2}}\int_{-m}^{+\infty}dx\frac{\sqrt{E_{F}^{2}-x^{2}}}{B}\theta(E_{F}-|x|)
\end{eqnarray}
when $E_{F}<m$,
\begin{eqnarray}
n &=& \frac{1}{(2\pi)^{2}}\int_{-E_{F}}^{+E_{F}}dx\frac{\sqrt{E_{F}^{2}-x^{2}}}{B}\nonumber\\
&=& \frac{1}{(2\pi)^{2}}\int_{-\pi/2}^{+\pi/2}d\theta \frac{E_{F}^{2}}{B}\cos^{2}\theta\nonumber\\
&=& \frac{E_{F}^{2}}{8\pi B}
\end{eqnarray}
when $E_{F}>m$,
\begin{eqnarray}
n &=& \frac{1}{(2\pi)^{2}}\int_{-\arcsin\frac{m}{E_{F}}}^{+\pi/2}d\theta  \frac{E_{F}^{2}}{B}\cos^{2}\theta \nonumber\\
&=& \frac{E_{F}^{2}}{8\pi B}[\frac{1}{2}+\frac{\arcsin(m/E_{F})}{\pi}+\frac{m}{\pi E_{F}}\sqrt{1-(m/E_{F})^{2}}].
\end{eqnarray}

\section{Derivation details of Eq.(\ref{34})}

In this appendix we provide the derivation of $\tilde{\omega}_{p,z}$ in the low doping regime [Eq.\ref{34})]. Taking advantage of the requirement that $\Gamma_{p,z}$ goes to 1 in the very low doping
regime where $\tilde{m}>>1$, $\tilde{B}<<1$, a combination of
Eq.(\ref{9}), Eq(\ref{10}), Eq.(\ref{23}) and Eq.(\ref{24}) in the main text leads to
\begin{eqnarray}
\Gamma_{p,z}^{-1}=\frac{1}{4}\tilde{\omega}_{p,z}^{2}[1-\frac{4}{3\pi\tilde{m}}+(\frac{1}{12}-\frac{4}{45\pi\tilde{m}^{3}})\tilde{\omega}_{p,z}^{2}]\rightarrow1,\label{C1}
\end{eqnarray}
Since $\tilde{m}>>1$ in the low doping regime, all terms containing $\tilde{m}$ can be neglected, and Eq.(\ref{C1}) reduces to
\begin{eqnarray}
\tilde{\omega}_{p,z}^{2}(1+\frac{1}{12}\tilde{\omega}_{p,z}^{2})\rightarrow4,\label{C2}
\end{eqnarray}
A straightforward calculation gives $\tilde{\omega}_{p,z}\rightarrow \tilde{\omega}_{z,c}=\sqrt{2(\sqrt{21}-3)}\approx1.78$.

To satisfy the self-consistent condition Eq.(\ref{self-c}) in the main text, both $\tilde{\omega}_{p,z}$ and
$\Gamma_{p,z}^{-1}$ should be kept to first order of $\tilde{B}/\alpha_{s}$. Suppose that $\tilde{\omega}_{p,z}$
takes the form of
\begin{eqnarray}
\tilde{\omega}_{p,z}=\tilde{\omega}_{z,c}+\beta\frac{\tilde{B}}{\alpha_{s}},\label{C3}
\end{eqnarray}
where $\beta$ is a constant to be determined. Bringing this form back to Eq.(\ref{self-c}), we find
\begin{eqnarray}
\Gamma_{p,z}^{-1}=1-2\tilde{\omega}_{z,c}^{2}\frac{\tilde{B}}{\alpha_{s}}+\mathcal{O}((\frac{\tilde{B}}{\alpha_{s}})^{2}),\label{C4}
\end{eqnarray}
As $\tilde{B}/\alpha_{s}<<1$, the second order term can be safely neglected. We can see that
$\Gamma_{p,z}^{-1}$ in Eq.(C4) satisfies the requirement given in Eq.(\ref{33}). To determine
the concrete value of $\beta$, we
bring Eq.(\ref{C3}) and Eq.(\ref{C4})
back into Eq.(\ref{C1}) and find that
\begin{eqnarray}
\beta=-\frac{24\tilde{\omega}_{z,c}}{7}.
\end{eqnarray}
Therefore, for $\tilde{B}/\alpha_{s}<<1$, the self-consistent solution to the first order of
$\tilde{B}/\alpha_{s}<<1$ is given as
\begin{eqnarray}
\tilde{\omega}_{p,z}=\tilde{\omega}_{z,c}-\frac{24\tilde{\omega}_{z,c}}{7}\frac{\tilde{B}}{\alpha_{s}}.
\end{eqnarray}
Similar procedures for $\omega_{p,\perp}$ gives, to the first order of $1/\alpha_{s}\tilde{m}$,
\begin{eqnarray}
&&\Gamma_{p,\perp}^{-1}=1-\tilde{\omega}_{\perp,c}^{2}\frac{1}{\alpha_{s}\tilde{m}},\nonumber\\
&&\tilde{\omega}_{p,\perp}=\tilde{\omega}_{\perp,c}-\frac{\tilde{\omega}_{\perp,c}^{3}}{2}\frac{1}{\alpha_{s}\tilde{m}}
\end{eqnarray}
with $\tilde{\omega}_{\perp,c}=2/\sqrt{1+(2\log(\tilde{\Lambda})/\pi\tilde{m})}\approx 2$.

\bibliography{dirac}

\end{document}